\documentclass{article}

\usepackage{PRIMEarxiv}

\usepackage[utf8]{inputenc} 
\usepackage[T1]{fontenc}    
\usepackage{hyperref}       
\usepackage{url}            
\usepackage{booktabs}       
\usepackage{amsfonts}       
\usepackage{nicefrac}       
\usepackage{microtype}      
\usepackage{lipsum}
\usepackage{fancyhdr}       
\usepackage{graphicx}       
\graphicspath{{media/}}     
\usepackage{makecell}
\usepackage{graphicx}%
\usepackage{multirow}%
\usepackage{amsmath,amssymb,amsfonts}%
\usepackage{amsthm}%
\usepackage{mathrsfs}%
\usepackage[title]{appendix}%
\usepackage{xcolor}%
\usepackage{textcomp}%
\usepackage{manyfoot}%
\usepackage{booktabs}%
\usepackage{algorithm}%
\usepackage{algorithmicx}%
\usepackage{algpseudocode}%
\usepackage{listings}%
\usepackage[numbers]{natbib}
\usepackage{makecell}

\title{A PLMs based Protein Retrieval Framework
\thanks{\textit{\underline{Citation}}: 
\textbf{Yuxuan Wu. A PLMs based Protein Retrieval Framework. Pages.... DOI:000000/11111.}} 
}

\author{
  Yuxuan Wu \\
  East China University of Science and Technology \\
  China \\
  Shanghai\\
  \texttt{wyuxuan@mail.ecust.edu.cn} \\
   \And
  Xiao Yi \\
   East China University of Science and Technology \\
  China \\
  Shanghai\\
  \texttt{Xiao\_Yis@outlook.com} \\
    \And
  Yang Tan\\
   East China University of Science and Technology \\
  China \\
  Shanghai\\
  \texttt{tyang@mail.ecust.edu.cn} \\
    \And
  Huiqun Yu\\
   East China University of Science and Technology \\
  China \\
  Shanghai\\
  \texttt{yhq@ecust.edu.cn} \\
    \And
  Guisheng Fan\\
   East China University of Science and Technology \\
  China \\
  Shanghai\\
  \texttt{gsfan@ecust.edu.cn} \\
      \And
  Gaowei Zheng\\
   East China University of Science and Technology \\
  China \\
  Shanghai\\
  \texttt{gaoweizheng@ecust.edu.cn} \\
}

\begin{document}
\maketitle

\begin{abstract}
Protein retrieval, which targets the deconstruction of the relationship between sequences, structures and functions, empowers the advancing of biology. Basic Local Alignment Search Tool (BLAST), a sequence-similarity-based algorithm,  has proved the efficiency of this field. Despite the existing tools for protein retrieval, they prioritize sequence similarity and probably overlook proteins that are dissimilar but share homology or functionality. In order to tackle this problem, we propose a novel protein retrieval framework that mitigates the bias towards sequence similarity. Our framework initiatively harnesses protein language models (PLMs) to embed protein sequences within a high-dimensional feature space, thereby enhancing the representation capacity for subsequent analysis. Subsequently,  an accelerated indexed vector database is constructed to facilitate expedited access and retrieval of dense vectors. Extensive experiments demonstrate that our framework can equally retrieve both similar and dissimilar proteins. Moreover, this approach enables the identification of proteins that conventional methods fail to uncover. This framework will effectively assist in protein mining and empower the development of biology.
\end{abstract}

\keywords{Protein retrieval \and Protein language model \and Deep learning \and Vector database}

\section{Introduction}
In the post-genomic era, proteins occupy a central stage of inquiry\cite{bib1}, catalyzing the advent of advanced retrieval techniques – including mass spectrometry, tank mass spectrometry (MS/MS), and high-performance liquid chromatography (HPLC)\cite{bib2,bib3} – designed to illuminate protein abundance and functional characteristics. The exponential growth of biological data in recent times has fueled progress in computational biology\cite{bib4}. Protein retrieval, a pivotal and intricate procedure in bioinformatics research, necessitates comprehensive assessment of diverse genomic variations: deletions, insertions, rearrangements, and mutations, among others. This process has evolved into a linchpin of contemporary biological investigations\cite{bib5}, underscoring its criticality in deciphering the complexity of proteomic landscapes.

Due to the fact that proteins exhibiting comparable primary structures are more likely found to share structural and functional similarities, many protein retrieval techniques have been proposed\cite{bib6,bib7,bib8}. Chief among these is primary sequence alignment, which revolves around identifying alignment methodologies that optimize sequence similarity. Notably, this alignment challenge is mathematically complex, classified as NP-complete\cite{daugelaite2013overview}, and commonly managed using scoring matrices that encapsulate amino acid substitution values for aligned pairs\cite{edgar2006multiple}. This has given rise to algorithms such as the Needleman-Wunsch global alignment for exhaustive comparisons\cite{bib6}, as well as local alignment approaches like Smith-Waterman\cite{bib7} and BLAST\cite{bib8} for pinpointing high-scoring segments indicative of functional resemblance amidst shorter alignments. While these strategies have significantly advanced protein retrieval capabilities, they inadvertently emphasize sequence similarity to the potential detriment of homologous yet sequence-divergent proteins\cite{bib9}, undermineing comprehensive understanding of proteins. An ideal retrieval mechanism ought to surpass the confines of basic sequence alignment, aspiring to a sophisticated grasp of protein attributes that encapsulates both structure and function.

To deal with the issues, a PLMs based protein retrieval framework, leveraging advancements in deep learning and language models, is introduced. PLMs have made significant progress in coevolution-based contact prediction\cite{bib13,bib14,bib15}, structure prediction\cite{bib16,bib17}, and unsupervised pretraining for contact prediction tasks, which means that PLMs have achieved a deep understanding of proteins to some extent. Firstly, to extract potential information from PLMs, we modify the models and use their encoders to obtain high-dimensional embeddings of protein sequences; Secondly, in the inference section, half-precision inference and pruning techniques are incorporated to improve the inference efficiency for large-scale databases; Thirdly, to accelerate the access and retrieval of dense vectors during retrieval, vector index tree (VPTree) and facebook AI similarity search (FAISS) are used to construct a high-dimensional embedded database. Moreover, local sensitive hash (LSH) is employed for dimensionality reduction and clustering of embeddings to facilitate distributed storage and parallel computation. Subsequently, similarity scores between encoded queries and database entries are compared to rank and extract the top-K matches. Extensive experiments suggest that our framework is able to avoid the local optimal trap caused by sequence similarity, and detect proteins ignored by conventional approaches with little performance compromise in accuracy and stability. Additionally, we examine the performance of diverse PLMs under this framework, underscoring the influence of the underlying PLM on retrieval efficacy.

The main contributions of this paper are summarized as follows:

\begin{itemize}
    \item We propose a protein retrieval framework based on PLMs, which provides a deep understanding of proteins through the model and overcomes the challenges of dense vector access and comparison. It deals with the retrieval unbalance caused by sequence similarity.  

    \item We introduce a benchmark for PLMs in retrieval field. It uses EC numbers as positive standards and is able to test the retrieval performance of various PLMs. Additionally, the benchmark is organized into a pipeline, which receives the embedding of PLM and comprehensively calculates retrieval metrics, thereby quickly and conveniently evaluating the retrieval performance of the model.
    
\end{itemize}

Section 2 discusses the preliminary content of the work related to protein retrieval methods and PLMs. Section 3 discusses the proposed protein retrieval framework and introduces the principles and methods of different parts within the framework. In Section 4, the experimental setup and multi angle experimental results are described, and the conclusion is drawn in Section 5.

\section{Preliminaries}

This section discusses the literature related to established protein retrieval methods. Additionally, it provides a brief overview of PLMs.

\subsection{Protein retrieval}\label{subsec1}

Protein retrieval has long since emerged as a vibrant arena of research, enabling substantial reductions in both the fiscal costs and operational intricacies inherent to protein target identification.

Table \ref{tab:mytable1} compiles an enumeration of established protein retrieval methodologies, categorizing them according to their foundational reliance on either primary or tertiary protein structures. These techniques encompass rule-based systems, characterized by user-defined combinations of metrics and thresholds, alongside machine learning (ML) and deep learning (DL). It is important to clarify that although DL is a type of ML method, it is separated from ML in this table because their feature extraction and model training processes may differ significantly.

\begin{table}[ht]
\centering
\caption{Protein retrieval methods}
\label{tab:mytable1}
\begin{tabular}{ccccc}
\hline
                                                           &                      & \multicolumn{3}{c}{Used Metric/Feature   Types} \\
Method                                                             & Mechanism & Sequence       & Structure       & Family       \\ \hline
MS/MS Experiment                                                & Rule     & N              & N               & Y            \\
Needleman-Wunsch                                               & Rule  & Y              & N               & N            \\
Smith-Waterman                                                 & Rule  & Y              & N               & N            \\
Basic Local Alignment Search   Tool(BLAST)                & ML       & Y              & N               & N            \\
Combinatorial Extension(CE)                                 & ML       & N              & Y               & N            \\
Combinatorial Extension for   Circular Permutations(CE-CP)  & ML       & N              & Y               & N            \\
Foldseek(TM-align)                                          & DL       & Y              & Y               & N            \\ \hline

\end{tabular}
\end{table}

\subsubsection{Primary sequence alignment}\label{subsubsec1}

Proteins' primary sequences are precisely constructed by connecting multiple amino acids in a linear chain in a specific order. Each amino acid molecule contains an amino group (NH2), a carboxyl group (COOH), and a unique side chain (R-group).

Sequence alignment algorithms strive to identify the optimal alignment between two sequences, which is an NP-complete problem, where  dynamic programming is often required. Based on this consideration, the most common one is the Needleman-Wunsch global alignment algorithm\cite{bib6}, which was devised to accommodate single nucleotide polymorphisms (SNPs), insertions, and deletions via a dynamic scoring system for residue pair alignments. This method requires maintaining a matrix based on the length of the sequence, and comparing the corresponding amino acids at each node. Progressing from this foundation, methods like Smith-Waterman\cite{bib7} were developed, combining global alignment with progressive strategies, refining sequence analysis through localized alignment optimizations.

\begin{figure}[ht]
    \centering
    \includegraphics[scale=0.11]{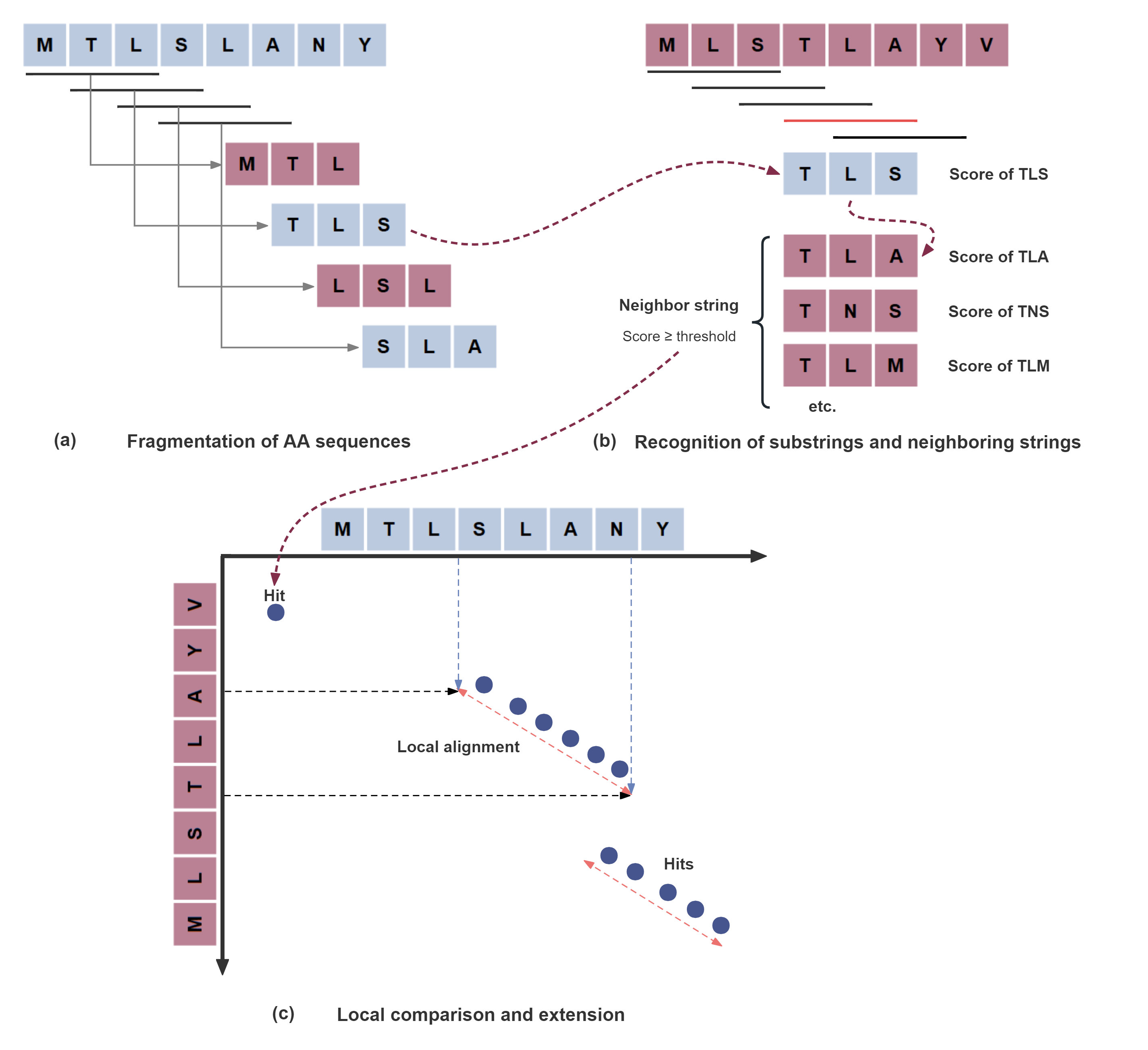}
    \caption{BLAST algorithm diagram}
    \label{mypic:03}
\end{figure}

To economize on alignment resources, the concept of local alignment was introduced, spurring the development of numerous alignment algorithms, notably including BLAST\cite{bib8}. Its central tenet revolves around fragmenting amino acid sequences and harnessing dynamic programming for flexible matching. Initially, an input is segmented into k-mers (Fig. \ref{mypic:03}(a)), with k commonly set to 3 for proteins and 11 for DNAs; shorter k-mers, while enhancing sensitivity, incur elevated computational demands. Thereafter, BLAST identifies matching segments in the complementary sequence exceeding a predefined score threshold, labeled as neighboring segments (Fig. \ref{mypic:03}(b)). These segments are queried against the database, and their detections, or 'hits', potentially cluster along diagonal paths indicative of high local similarity (Fig. \ref{mypic:03}(c)). The process culminates in bidirectional extension of hits until the alignment score dips beneath a set criterion, delineating the boundaries of homologous regions.

\subsubsection{Tertiary structure alignment method}\label{subsubsec2}

Another alignment avenue incorporates proteins' tertiary structure information, embodied in algorithms such as the Combinational Extension (CE)\cite{bib10} and its variant tailored for circular permutations (CE-CP)\cite{bib11}. These methodologies integrate dynamic programming with sliding window tactics to discern optimal local homologies.

Despite offering enhanced accuracy, structural alignment poses a considerable computational time challenge compared to sequence-based methods. To mitigate this, Michel van Kempen and colleagues advocated for Foldseek(TM-align)\cite{bib12}, an innovative strategy that transforms protein structural data into a sequence-like format, thereby bridging the gap between structural and sequence alignment paradigms. Leveraging deep learning and a pretrained structural alphabet, Foldseek generates sequences  from structural data, which serve as the basis for comparing and identifying similarities across protein architectures. Nonetheless, these advancements are contingent upon the availability of known protein structures, a limitation exacerbated by the daily influx of novel protein yet to have their structures resolved.

\subsection{Protein language models}\label{subsec2}

Deep learning is a branch of machine learning that aims to simulate the brain's neural network to learn and understand complex patterns. The innovation lies in multi-layered neural networks that learn complex features from data, handling nonlinear tasks such as image and speech recognition, and natural language processing (NLP). Language models are center of NLP, which decode language structure for machine understanding and generation. Protein language models (PLMs) are specialized adaptations trained to comprehend biological data intricacies.

Significant strides have been witnessed in coevolution-driven contact prediction\cite{bib13,bib14,bib15}, structure prediction methodologies\cite{bib16,bib17}, and unsupervised pre-training for contact prediction tasks. Notably, AlphaFold2 stands as a state-of-the-art case in structure prediction, leveraging primary sequences to forecast intricate tertiary structures with unprecedented precision, as evidenced by its performance in CASP14\cite{bib18}. The achievements highlighted reveal PLMs' capacity for attaining an extensive understanding of proteins, facilitating precise deduction of their classifications, abundance levels, and functional aspects from either primary or tertiary information.

\section{Proposed approach}\label{sec3}

\begin{figure}[ht]
    \centering
    \includegraphics[scale=0.71]{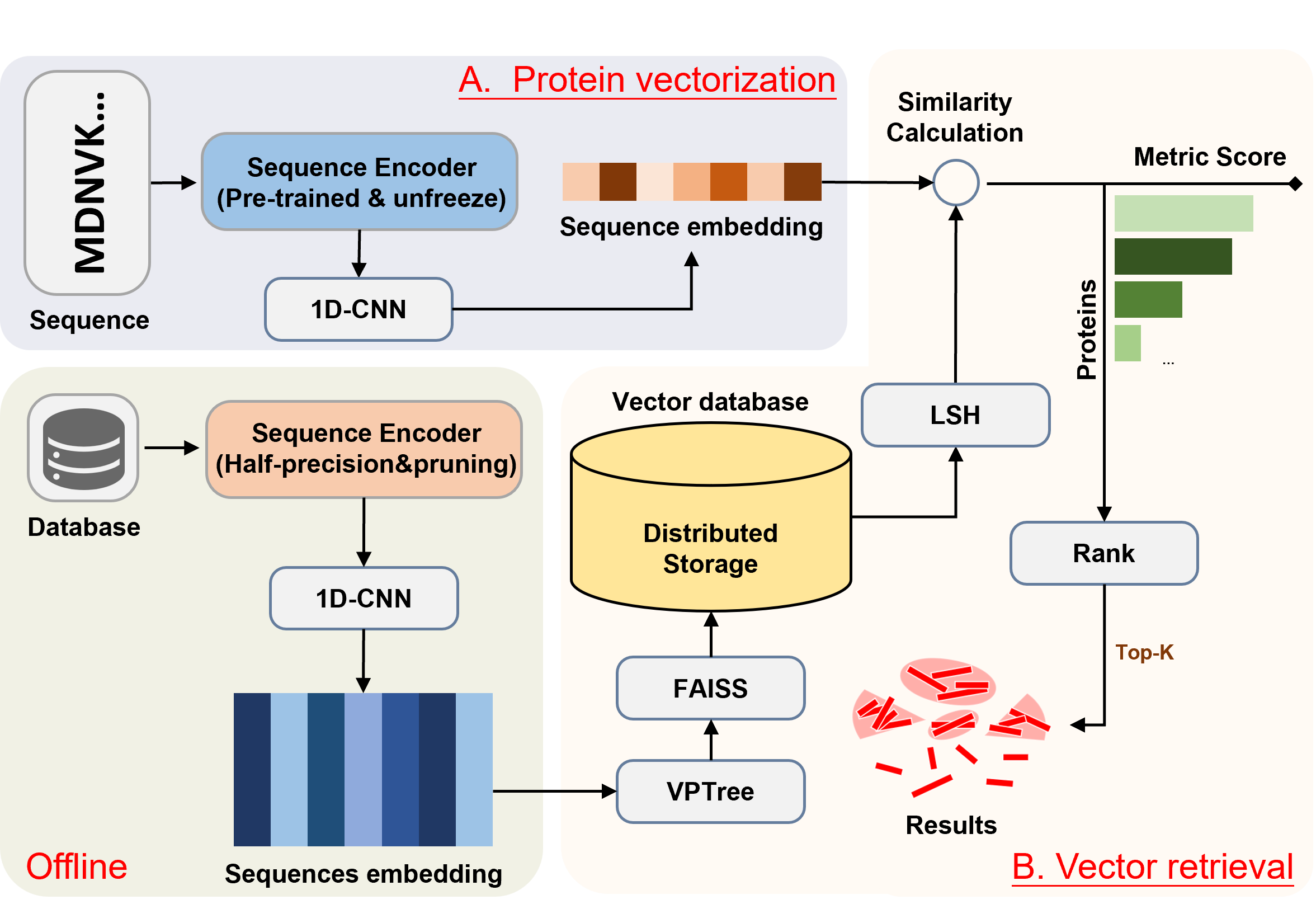}
    \caption{A PLMs based Protein Retrieval Framework}
    \label{mypic:06}
\end{figure}

This section delineates the proposed PLMs based protein retrieval framework, which is comprised of protein vectorization and vector retrievel module, as shown in Fig. \ref{mypic:06}. Commencing with the protein vectorization module, it consists of encoders of query sequences and candidate sequences (sequences from a very large database). The vector retrieval module is responsible for storing and accessing dense vectors, as well as vectors comparison.

Preparation before retrieval is transforming the sequence database into embeddings using candidate sequences encoders, which is a massive sequence of reasoning process, therefore semi precision reasoning and pruning techniques are used to accelerate this process. After obtaining dense embeddings, VPTree and FAISS are used to establish indexes and block data for reasonable storage in vector retrieval module. 

For a retrieval, query sequences encoders of protein vectorization module encode query sequence as an embedding. When it enters the vector retrieval module, it will be compared with each protein in the vector database. To accelerate the access, locality sensitive hashing, which is an approximate nearest neighbor fast search technique for massive high-dimensional data, is used. Subsequently, the framework sorts the database according to the calculated similarity scores and selected the top-k entries as the core retrieval results, ensuring relevance and precision in our retrieval process.

Details of protein vectorization and vector retrieval module are introduced as follow.

\subsection{Protein vectorization}

To construct the retrieval dataset, protein sequences are required to be vectorized. The primary protein sequences consist of the amino acid, wherein each amino acid is denoted by the corresponding letters in Table \ref{tab:mytable222}. Therefore, a primary sequence is defined as $P=p_1,p_2,\ldots,p_i,\ldots,p_T$, where $p_i (i\in\mathbb{R})$ is the amino acid character corresponding to the i-th position. 

\begin{table}[ht]
\centering
\caption{Names and symbol representations of 20 common amino acids}
\label{tab:mytable222}
\begin{tabular}{cccccc}
\hline
Name                                 & 3-letter code & 1-letter code & Name                                 & 3-letter code & 1-letter code \\ \hline
{\color[HTML]{2C2C36} Alanine}       & Ala               & A                  & {\color[HTML]{2C2C36} Leucine}       & Leu               & L                  \\
{\color[HTML]{2C2C36} Arginine}      & Arg               & R                  & {\color[HTML]{2C2C36} Lysine}        & Lys               & K                  \\
{\color[HTML]{2C2C36} Asparagine}    & Asn               & N                  & {\color[HTML]{2C2C36} Methionine}    & Met               & M                  \\
{\color[HTML]{2C2C36} Aspartic acid} & Asp               & D                  & {\color[HTML]{2C2C36} Phenylalanine} & Phe               & F                  \\
{\color[HTML]{2C2C36} Cysteine}      & Cys               & C                  & {\color[HTML]{2C2C36} Proline}       & Pro               & P                  \\
{\color[HTML]{2C2C36} Glutamine}     & Gln               & Q                  & {\color[HTML]{2C2C36} Serine}        & Ser               & S                  \\
{\color[HTML]{2C2C36} Glutamic acid} & Glu               & E                  & {\color[HTML]{2C2C36} Threonine}     & Thr               & T                  \\
{\color[HTML]{2C2C36} Glycine}       & Gly               & G                  & {\color[HTML]{2C2C36} Tryptophan}    & Trp               & W                  \\
{\color[HTML]{2C2C36} Histidine}     & His               & H                  & {\color[HTML]{2C2C36} Tyrosine}      & Tyr               & Y                  \\
{\color[HTML]{2C2C36} Isoleucine}    & Ile               & I                  & {\color[HTML]{2C2C36} Valine}        & Val               & V                  \\ \hline
\end{tabular}
\end{table}

The initial step of  vectorization is transforming primary sequences into 1-hot encodings, which serves as an encoding format readily comprehensible to the model. Nonetheless, this encoding method has limitations for its dimensionality scaling exponentially with sequence length. To mitigate the computational burden imposed by excessively lengthy sequences, certain PLMs impose constraints on input 1-hot encoding length, thereby necessitating the pivotal step of padding or truncating. PLMs adopting absolute positional encoding are restricted to a 1-hot encoding length of 1024, while those implementing rotary position encoding are limited to a cap of 7002. In batch processing, sequences not meeting the established length criteria are adjusted by appending padding elements [PAD] to attain the necessary identifier length. However, this will lead to output contaminated with noise. To deal with this, special tokens [CLS] and [SEP] are respectively incorporated as initiation and termination markers for truncating of outputs.

\begin{equation}
Attention(Q, K, V) = softmax(\frac{QK^T}{\sqrt{d_k}})V
\end{equation}

\begin{equation}
FFN(x) = max(0, xW_1 + b_1)W_2 + b_2
\end{equation}

\begin{equation}
LayerNorm(x + Sublayer(x))
\end{equation}

In the subsequent phase, the encoding layer harnesses a multi-head self-attention mechanism, depicted in equation 1, to discern the interdependencies among diverse positions within the inputs. $Q,K,V$ are linear transformation of input and $d_k$ is the dimension of tensors. Following this, a fully-connected feedforward neural network, grounded in equation 2, is employed, where $W_1,W_2$ are weight coefficients, and $b_1,b_2$ are paranoid coefficient. This network executes linear and nonlinear transformations on individual positional representations, thereby enhancing the richness of feature embeddings. Following embedding and truncation procedures, individual protein sequences are transformed into a digital matrix representation, serving as input to a 1D-CNN for feature extraction through pooling operations.

\begin{algorithm}
\caption{Protein vectorization}\label{algo1}
\begin{algorithmic}
\Require $P=p_1,p_2,\ldots,p_i,\ldots,p_T (i\in\mathbb{R})$
\Ensure $embeddings\;of\;P$
\State $d := length(P)$ 
\State $flag := 0$
\If{$d < input\_dimension$}\label{algln2}
        \State $P \Leftarrow [pad]$
        \State $flag := 1$
\Else
        \State $P := Truncation(P)$
\EndIf
\State $P \Leftarrow [cls]$
\State $P \Leftarrow [sep]$
\State $N \:= length(P)$
\While{$N \neq 0$}
        \State $X=one\_hot\_encoding(P[N])$
        \State $y=encoders\_layer(X)$
        \State $Y \Leftarrow y$        
        \State $N := N - 1$
\EndWhile
\If{ $flag = 0$ }\label{algln3}
        \State $embeddings := 1D-CNN(Y)$
\Else
        \State $Y := Truncation(Y)$
        \State $embeddings := 1D-CNN(Y)$
\EndIf

\end{algorithmic}
\end{algorithm}

\begin{figure}[ht]
    \centering
    \label{mypic:10}
    \includegraphics[scale=0.69]{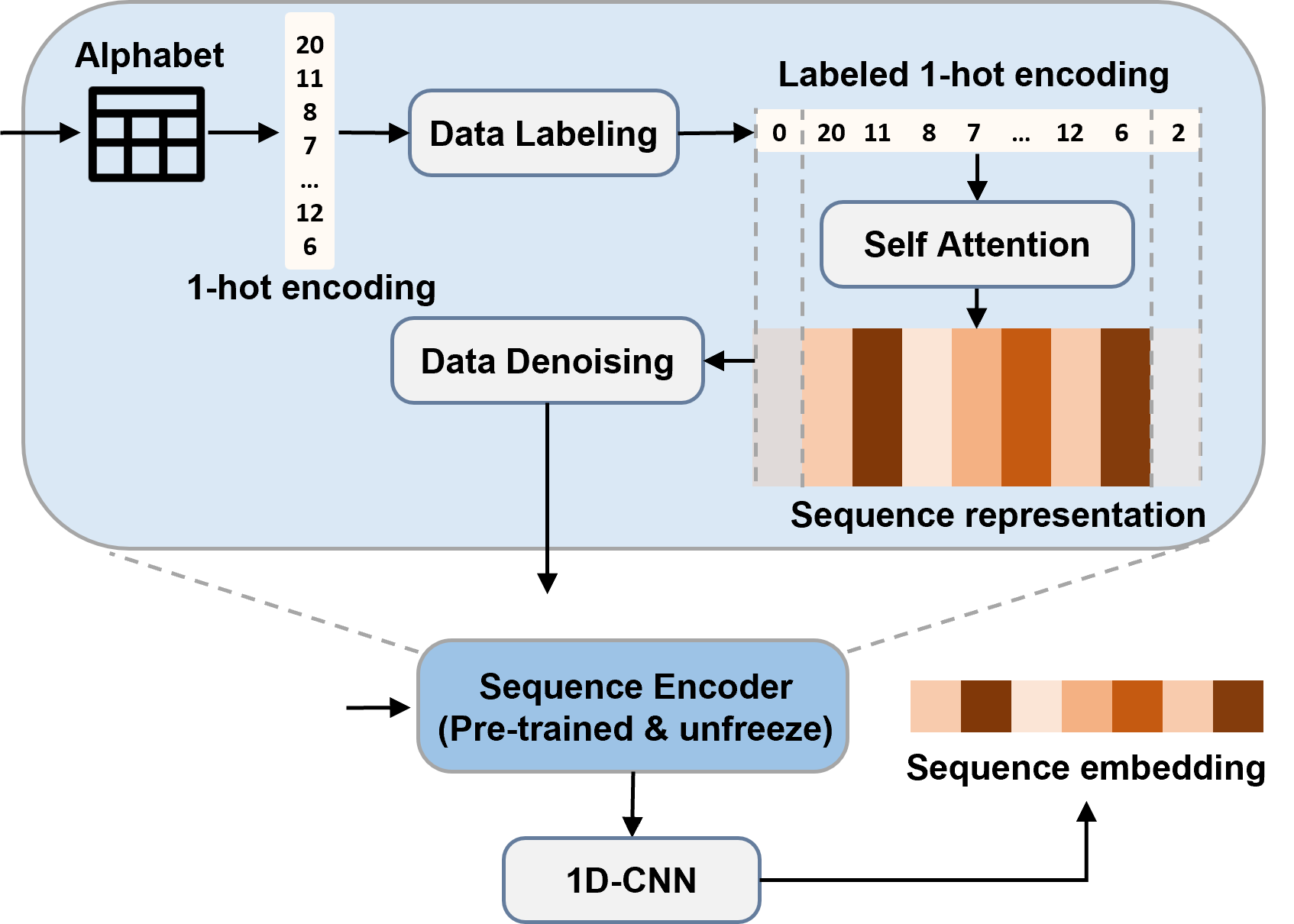}
    \caption{Embedding of protein sequence}
\end{figure} 

\subsection{Vector retrieval}

\begin{figure}[ht]
    \centering
    \includegraphics[scale=0.27]{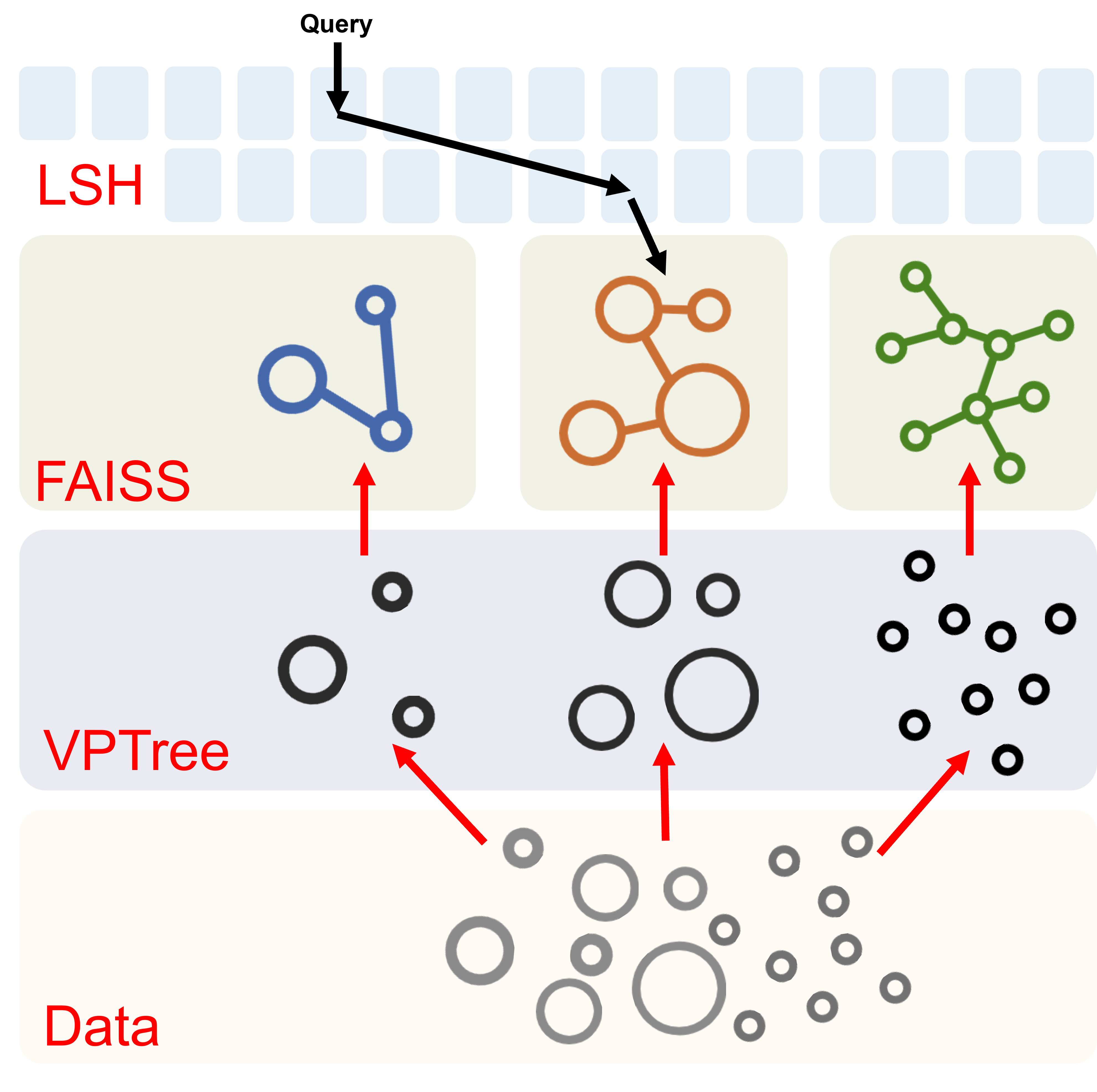}
    \caption{Multi-layer acceleration structure}
    \label{mypic:index}
\end{figure}

The primary challenge confronted by the vector retrieval module pertains to the efficient storage and retrieval of dense vectors. To tackle the aforementioned challenges, the employment of VPTree index as the foundational indexing mechanism for dense vector datasets is deemed imperative. This sophisticated algorithm ingeniously erects a hierarchical data partitioning framework, wherein a succession of strategic center points, or vantage points, are meticulously chosen to segment the data universe. It facilitates an efficient and systematic delineation of the vector space, thereby significantly expediting the process of proximity search whilst preserving the integrity and granularity of the underlying data structure. In addition, using FAISS as the upper level index can significantly improve search speed while maintaining high accuracy. In order to further improve retrieval speed, we use local sensitive hashing technology to construct a hash table with multiple layers of indexes. As shown in Fig.\ref{mypic:index}, the multi-layer acceleration structure solves the problem of dense vector retrieval and provides acceleration support for the vector retrieval module.

Upon constructing retrieval and FAISS indexes, it is necessary to provide an appropriate similarity metric. Given the absence of comprehensive studies dictating the most suitable measure for this endeavor, meticulous consideration is essential, as varied metrics gauge similarity from disparate angles. To effectively retrieve proteins akin to the query within the latent space, the choice of a fitting similarity criterion is paramount.

Commonly adopted benchmarks for evaluating latent space representations encompass the inner product (IP), euclidean distance (L2), and cosine similarity. These metrics find broad application across domains such as computer vision, natural language processing, and recommendation systems. Specifically, the inner product computes the dot product of vectors, quantifying latent space similarity where a larger product signifies greater proximity. Conversely, the Euclidean distance, measured as the direct line-length between vectors, denotes dissimilarity, with shorter distances indicating higher similarity. Cosine similarity, assessing vector alignment by calculating the cosine of the angle between them, spans from -1 to 1, nearing 1 for aligned vectors, -1 for opposing directions, and tending towards 0 for orthogonal vectors.

Furthermore, recognizing the mathematical equivalence of cosine similarity to the normalized inner product by the L2 norm (as per equations 4 and 5), we also examine the euclidean distance of normalized embeddings, referred to here as norm\_L2. Collectively, our investigation encompasses a quartet of distinct similarity indices, meticulously selected to comprehensively evaluate vector similarity in the context of protein retrieval.

\begin{equation}
\begin{split}
norm\_IP = IP(\ldots,{\frac{x_i}{\sqrt{x_1^2+x_2^2+\ldots+x_n^2}}},\ldots,{\frac{y_i}{\sqrt{y_1^2+y_2^2+\ldots+y_n^2}},\ldots}) \label{05}
\end{split}
\end{equation}

\begin{equation}
\begin{split}
COSINE=\frac{x_1*y_1+x_2*y_2+\ldots+x_n*y_n}{\sqrt{x_1^2+x_2^2+\ldots+x_n^2}*\sqrt{y_1^2+y_2^2+\ldots+y_n^2}} \label{06}
\end{split}
\end{equation}

\section{Experiments and results}\label{sec4}

\subsection{Notation and setup}\label{subsec4}

To benchmark the retrieval effectiveness of diverse methodologies and frameworks, we anchor our evaluation on the esteemed UniProtKB/SwissProt protein sequence repository, employing the Enzyme Commission (EC) number as a positive discriminant. UniProtKB\cite{bib31} is a widely used protein database maintained by the SIB (Swiss Institute of Bioinformatics) in collaboration with the EBI (European Bioinformatics Institute). It can be broadly categorized into UniProtKB/TrEMBL and UniProtKB/SwissProt, providing extensive and detailed protein sequence and annotation information. Proteins in UniProtKB/SwissProt have undergone rigorous experimental validation and verification. Therefore, SwissProt is renowned for its precision, high quality, and manual curation, making it one of the earliest established protein databases in the world. It contains protein sequences and related information from various biological species, including annotations on protein functions, structures, interactions, metabolic pathways, disease associations, and more.

The EC number, a functional taxonomy marker embedded in the SwissProt database, imparts a systematic classification to enzymes in accordance with the biochemical reactions they catalyze. Endorsed by the International Union of Biochemistry and Molecular Biology (IUBMB), this identifier adopts a structured format of 'EC' followed by a quartet of digits separated by periods, each conveying hierarchical information about the enzyme's activity. Specifically, the initial digit denotes the enzyme class, the subsequent digit identifies the subclass, the third digit further refines the category, and the terminal digit assigns a unique identifier within that subclass. This differentiation underscores the EC number's superiority over familial labels as an assessment criterion for retrieval systems, since it not only distinguishes between enzymes based on their precise biochemical actions but also reflects subtle variations in functionality.

Proteins typically have more than one EC number label. We define the EC number of each protein as  {$EC=\{ec_1,ec_2,\ldots,ec_m$\}}, where {$ec_i$} is a label in the form of a.b.c.d ( {$a, b, c, d \in R^+$} ).

To gauge retrieval precision, we adopt a four-tier matching schema for protein sequences. Level 4 denotes an exact match where EC numbers fully overlap ({$EC_\alpha \cap EC_\beta \neq \emptyset$}), signifying shared enzymatic functions. Level 3 is assigned when commonality is confined to the first three EC number digits, reflecting broader categorical agreement. Levels 2 and 1 further stratify decreasing levels of similarity based on EC prefixes, providing a comprehensive yet concise grading of retrieval accuracy.

\begin{figure}[htbp]
    \centering
    \label{mypic:07}
    \includegraphics[scale=0.65]{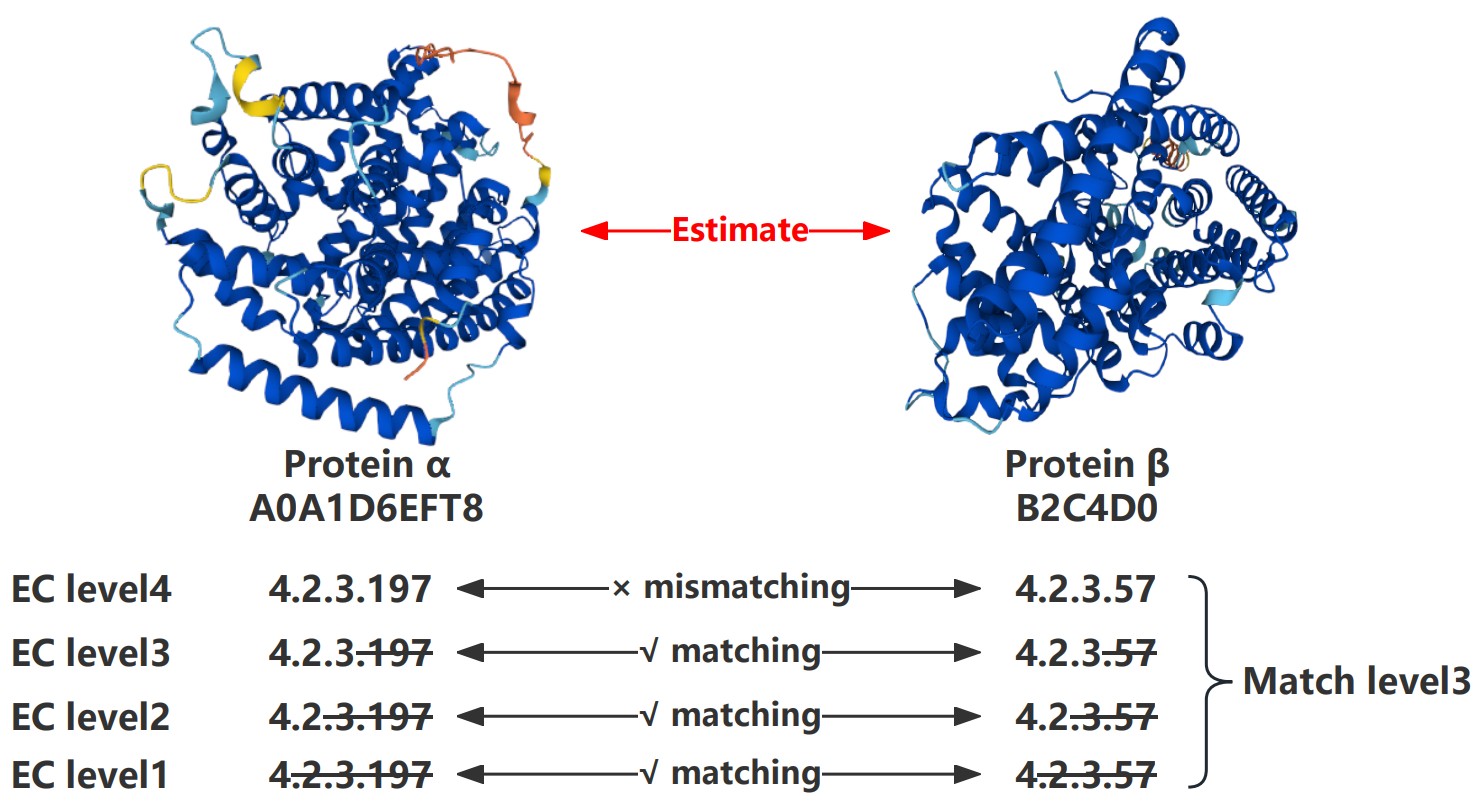}
    \caption{Protein retrieval multi-level matching diagram: In the diagram, protein {$\alpha$} refers to A0A1D6EFT8 with an EC number of 4.2.3.197, while protein {$\beta$} refers to B2C4D0 with an EC number of 4.2.3.57. The degree of matching between protein {$\alpha$} and protein {$\beta$} is assessed at four levels. Firstly, a level 4 evaluation is performed, and the intersection between {4.2.3.197} and {4.2.3.57} is empty, indicating no level 4 match. Next, a level 3 evaluation is conducted, and the intersection between {4.2.3} and {4.2.3} is not empty, indicating a level 3 match between protein {$\alpha$} and protein {$\beta$}. Similarly, protein {$\alpha$} and protein {$\beta$} also pass the level 2 and level 1 evaluations. Based on the above information, we conclude that the degree of matching between protein A0A1D6EFT8 and protein B2C4D0 is at level 3.}
\end{figure} 

The benchmark focuses on a subset of SwissProt, adorned with explicit EC labels, totally approximately 236,306 entries. Subsequently, 964 proteins with unique EC labels and crystal tertiary structure are randomly selected as queries. Notably, to address structural gaps, AlphaFold2 is enlisted to predict the missing tertiary structures for the dataset.

In addition, a pipeline based on the benchmark is established to evaluate the effectiveness of PLMs. The pipeline accepts the embeddings extracted from the subset, constructs a vector database, and performs retrieval on multiple similarity metrics. Then, four-level matching analysis ensues, ensuring a granular assessment of retrieval performance across varying degrees of functional alignment.

The source code for BLAST, Foldseek (TM-align), and the pipeline can be obtained from the following link: 

https://github.com/ncbi/blast\_plus\_docs

https://github.com/steineggerlab/foldseek 

https://github.com/ginnm/ProteinMiningEvaluator/tree/wyx

The experiment is conducted entirely on a platform consisting of RTX 4090 and 13th Gen Intel (R) Core (TM) i9-13900KF.

\subsection{Experimental Setup}\label{subsec5}

\textbf{Datasets} Evaluation is based on SwissProt\cite{bib32} protein database, which is dedicated to providing high-quality annotations, minimal redundancy, and high integration with other databases. It contains approximately 570,830 protein entries. A subset of 236,306 proteins with clear EC numbers is formed as retrieval dataset. From these, we meticulously select 964 proteins with explicit EC numbers and crystal tertiary structure as queries, which serve as probes in retrieval.  A comprehensive experiment in 964 retrieval scenarios is gained to evaluate the average performance of the framework. 

\textbf{Protein language models}  We conduct research and identify advanced and effective PLMs and methodologies. The most widely recognized model currently is esm1b\_t33\_650M\_UR50S, a Transformer-based protein language model capable of training on protein sequence data without the labeled supervision. This model is based on the RoBERTa\cite{bib19} architecture and training procedure, incorporating pre-activation layer normalization. It utilizes the Uniref50 2018\_03 protein sequence database for unsupervised Masked Language Modeling (MLM), where the model is trained to predict amino acids from the surrounding sequence context. Rao\cite{bib20} and Meier\cite{bib21} demonstrated the effectiveness of the model on various downstream tasks in 2020 and 2021, respectively. Additionally, we choose esm2\_t33\_650M\_UR50D and esm2\_t36\_3B\_UR50D\cite{bib22}. They are based on esm1b\_t33\_650M\_UR50S and optimizes absolute position encoding to rotated position encoding. The latter incorporating additional layers and attention heads on the former, further enhancing the performance. Furthermore, we evaluate esm\_msa1b\_t12\_100M\_UR50S, the variant with the addition of a column attention mechanism. It was subsequently applied by Bo Chen and Ziwei Xie to develop ColAttn\cite{bib23} in 2022. ColAttn evaluates the similarity relationships between multiple co-evolution information groups using esm\_msa1b\_t12\_100M\_UR50S and successfully enhances the prediction accuracy of AlphaFold-Multimer for complex proteins by integrating similar proteins. 

Considering the commendable performance of the T5 model in natural language processing, we test the prot\_t5\_xl\_uniref50\cite{bib24} (pre-trained in a self-supervised manner on the Uniref50 protein database, derived from the T5-3B). In contrast to the original T5-3B model, the prot\_t5\_xl\_uniref50 does not incorporate the span denoising objective used in the original T5-3B model. Instead, it employs an MLM denoising objective similar to Bert for pretraining. Similarly, we assess the classic autoregressive model ProtGPT2\cite{bib25}, which is based on the GPT2 Transformer architecture. It consists of 36 layers with a model dimension of 1280 and a total of 7.38 billion parameters. This decoder-transformer model is pre-trained in a self-supervised manner using the causal modeling objective on the UniRef50 2021\_04 protein sequence database. ProtGPT2 is specifically trained using the causal modeling objective to predict the next token in the sequence.

Structure-based models hold significant importance alongside sequence-based models. We evaluate the classic esm\_if1\_gvp4\_t16\_142M\_UR50\cite{bib26}, consisting of an invariant geometric input processing layer and a sequence-to-sequence Transformer. It is based on the prediction of 12 million protein structures by AlphaFold2. It is trained using scope masking to accommodate missing backbone coordinates, making it capable of predicting sequences of partially shielded structures.  We also choose MIF-ST\cite{bib27} (Masked Inverse Folding with Sequence Transfer) proposed by Kevin K. Yang et al. , which is based on the inverse folding concept. It features the ByteNet encoder architecture and pre-trains based on the Uniref50 2020\_05 dataset. In consideration of the valuable information embedded in protein structure sequences, the authors employed GNN networks for protein structure pre-training. They introduced the Masked Inverse Folding (MIF) task to pre-train the structure by utilizing masked amino acid sequence.

\textbf{Evaluation metrics} Our evaluation centers on the percent identity, a pivotal metric for revealing essential differences between the framework and traditional methods. Moreover, hit rate and true positive hits up to first false positive (TP hits up to the 1st FP) are considered to show the retrieval accuracy and stability. Ultimately, a protein mining report in 2023 is acted as further validation.

\subsection{Result}\label{subsec6}

\subsubsection{Percent Identity}\label{subsubsec5}

Protein sequence similarity is closely associated with functional similarity to a large extent. By comparing the similarity between protein sequences, it is possible to infer that they may have similar functional features. Proteins with highly similar sequences typically exhibit similar biological functions and structural domains, providing important clues for proteins with unknown functions. However, various protein retrieval methods have their own retrieval biases, and it is worth studying and evaluating how these biases are reflected in the Percent Identity metric.

In the evaluation of the Percent Identity metric, we conduct an in-depth analysis of the retrieval results of esm1b\_t33\_650M\_UR50S\_L2 and the BLAST method. We find that the protein retrieval method based on esm1b\_t33\_650M\_UR50S\_L2 held significant research value, in contrast to the BLAST method. We randomly select the case A0A0C5Q4Y6 from the 964 cases in this benchmark and analyze the direct differences of the positive samples in esm1b\_t33\_650M\_UR50S\_L2 and BLAST. We demonstrate the differences at different Topk levels through a Venn diagram.

\begin{figure}[htbp]
    \centering
    \label{mypic:14}
    \includegraphics[scale=0.25]{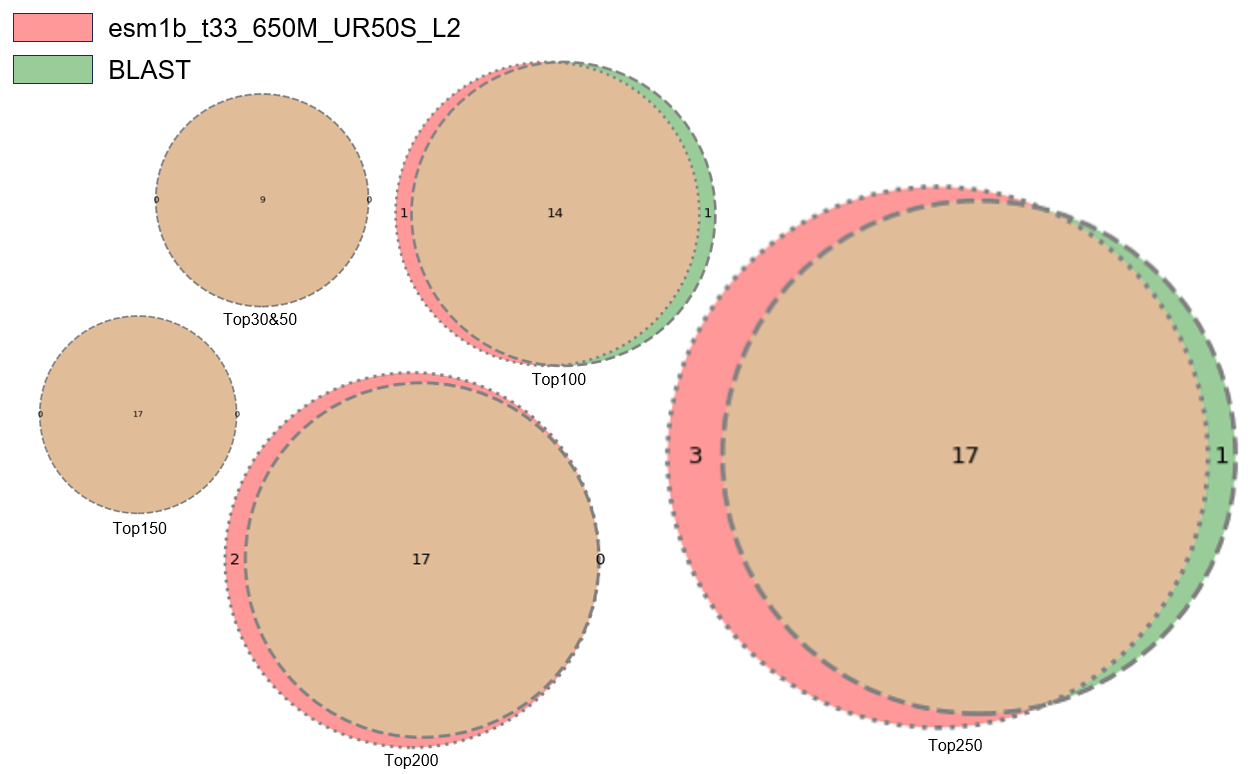}
    \caption{Venn comparison}
\end{figure} 

\begin{figure}[htbp]
    \centering
    \includegraphics[scale=0.6]{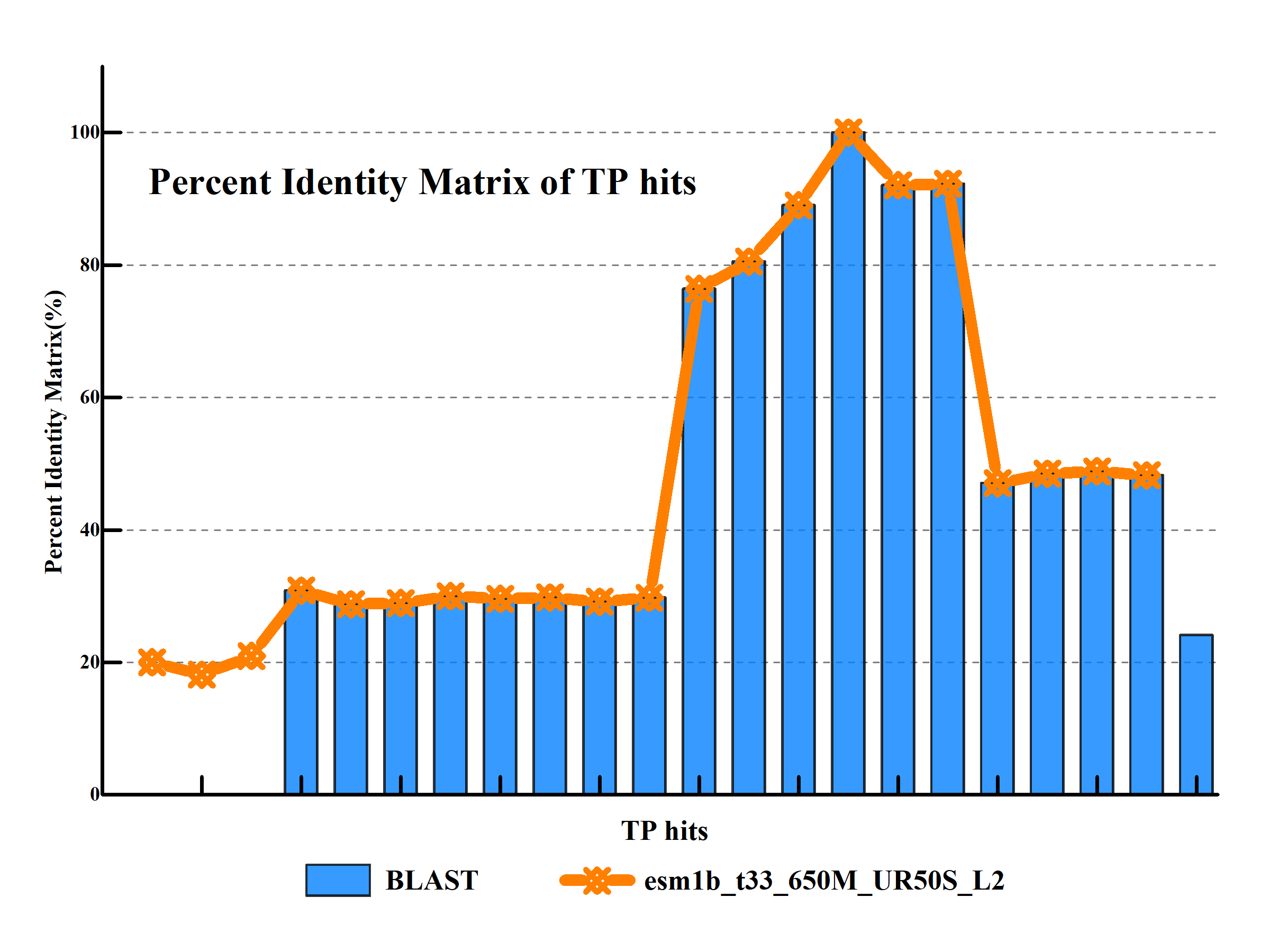}
    \caption{Percent Identity Matrix of TP hits}
    \label{mypic:11}
\end{figure}

We find that overall, the esm1b\_t33\_650M\_UR50S\_L2 could retrieve more positive samples than BLAST. In the cases of Top 30, 50, and 150, the positive samples retrieved by esm1b\_t33\_650M\_UR50S\_L2 completely include the retrieval results of BLAST. In the cases of Top 100 and Top 200, BLAST has only one unique positive sample, while esm1b\_t33\_650M\_UR50S\_L2 has 1 and 2 unique positive samples, respectively. In the case of Top 250, the esm1b\_t33\_650M\_UR50S\_L2 even has 3 unique positive samples. There must be intrinsic unknown factors at play within this. Therefore, we conduct a comprehensive analysis of all positive results from esm1b\_t33\_650M\_UR50S\_L2 and BLAST. The situation is shown in Figure \ref{mypic:11}.

The detailed comparison of all positive samples with protein A0A0C5Q4Y6 is shown in Table \ref{tab:mytable4}, including 3 unique positive samples from esm1b\_t33\_650M\_UR50S\_L2, 1 unique positive sample from BLAST, and 17 positive samples in common.

\begin{table}[ht]
\centering
\caption{PI Matrix of retrieval results for A0A0C5Q4Y6}
\label{tab:mytable4}
\begin{tabular}{ccc}
\hline
Protein    & esm1b\_t33\_650M\_UR50S\_L2 & BLAST     \\ \hline
Q9M066     & 20\%                        & Not Found \\
Q94IA6     & 18.12\%                     & Not Found \\
I1HL09     & 20.97\%                     & Not Found \\ \hline
P0DO35     & \multicolumn{2}{c}{30.83\%}             \\
P0DO42     & \multicolumn{2}{c}{28.75\%}             \\
P0DO41     & \multicolumn{2}{c}{28.96\%}             \\
P0DO40     & \multicolumn{2}{c}{30.00\%}             \\
P0DO38     & \multicolumn{2}{c}{29.58\%}             \\
P0DO39     & \multicolumn{2}{c}{29.79\%}             \\
O49394     & \multicolumn{2}{c}{29.16\%}             \\
Q9SZ46     & \multicolumn{2}{c}{29.77\%}             \\
A0A1Z3GBS4 & \multicolumn{2}{c}{76.42\%}             \\
S4UX02     & \multicolumn{2}{c}{80.53\%}             \\
A0A0Y0GRS3 & \multicolumn{2}{c}{89.05\%}             \\
A0A0C5Q4Y6 & \multicolumn{2}{c}{Itself}              \\
A0A0C5QRZ2 & \multicolumn{2}{c}{92.07\%}             \\
A0A0S1TP26 & \multicolumn{2}{c}{92.27\%}             \\
A0A1D8QMD2 & \multicolumn{2}{c}{47.05\%}             \\
A0A1D8QMG4 & \multicolumn{2}{c}{48.47\%}             \\
A0A1D8QMD1 & \multicolumn{2}{c}{48.86\%}             \\
A0A0S1TPC7 & \multicolumn{2}{c}{48.23\%}             \\ \hline
P51589     & Not Found                   & 24.10\%   \\ \hline
\end{tabular}

\end{table}

We conduct sequence alignment of all positive results with A0A0C5Q4Y6 and determine the percent identity between the positive results and A0A0C5Q4Y6. Then, we visualize the results and found a slight pattern in the differences among the positive results. The percent identity between the unique positive results from esm1b\_t33\_650M\_UR50S\_L2 and A0A0C5Q4Y6 is approximately 20\% or lower, with the lowest percent identity reaching 18.12\%. On the other hand, the percent identity of the unique positive result from BLAST is 24.10\%, while the lowest percent identity among the common positive results is only 28.75\%.

\begin{figure}[htbp]
    \centering
    \includegraphics[scale=0.3]{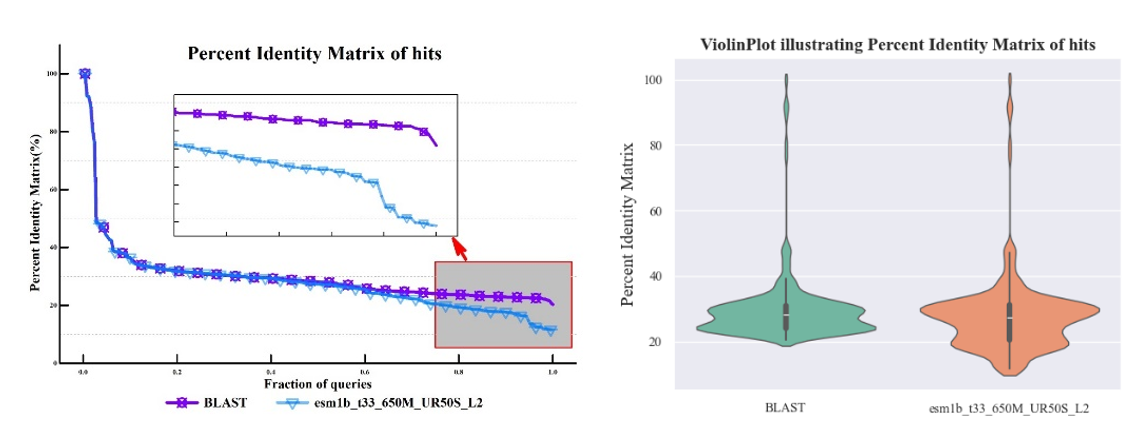}
    \caption{Percent Identity Matrix of hits}
    \label{mypic:12}
\end{figure}

To further validate this pattern, we conduct sequence alignment and sorting visualization of all retrieval results (including both positive and false positive samples) from both methods. The lowest percent identity among the retrieval results from BLAST is 20.4\%, while the lowest percent identity among the retrieval results from esm1b\_t33\_650M\_UR50S\_L2 was 11.6\%. The results are shown in Figure \ref{mypic:12}.

To observe the differences between esm1b\_t33\_650M\_UR50S\_L2 and BLAST more intuitively, we sort the retrieval results by rank instead of percent identity and visualize them using line charts.

\begin{figure}[htbp]
    \centering
    \includegraphics[scale=0.7]{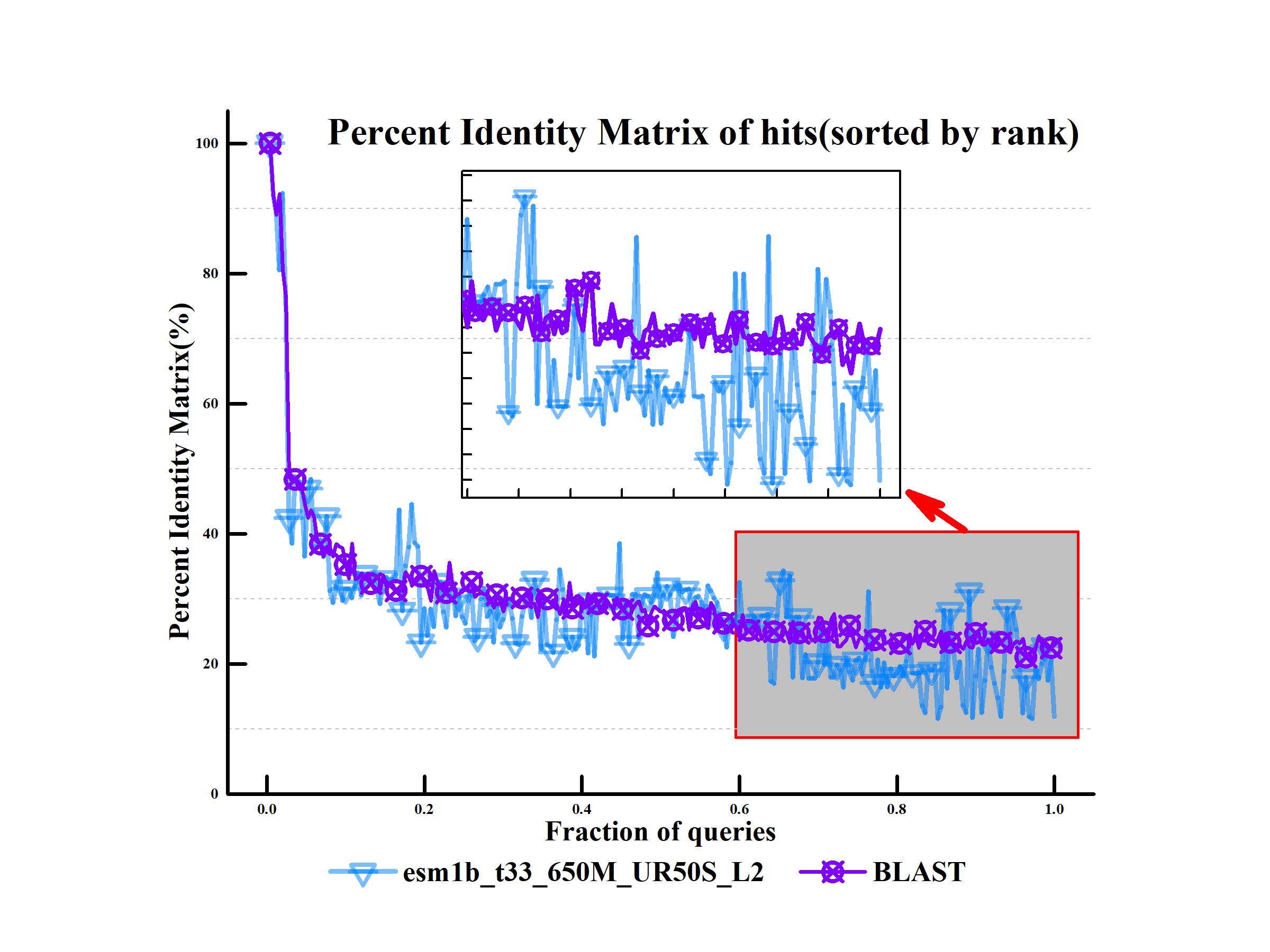}
    \caption{Percent Identity Matrix of hits(sorted by rank)}
    \label{mypic:13}
\end{figure}

From Figure \ref{mypic:13}, it could be observed more intuitively that the retrieval results from BLAST show a decreasing trend in percent identity as the rank increases. This suggests that BLAST retrieval is more likely to capture samples with high percent identity rather than being solely based on a comprehensive understanding of proteins for retrieval. On the other hand, the retrieval method based on esm1b\_t33\_650M\_UR50S\_L2 indicates that samples with low percent identity could also rank higher in the retrieval, exhibiting a fluctuating trend similar to noise. There is no trend of decreasing percent identity as the rank increases. This suggests that in the PLMs based retrieval framework, samples with low percent identity are not discarded but are equally included in the retrieval process. Percent identity is not the core criterion in this framework.

We then randomly select several cases for analysis, further confirming our hypothesis. We find that the retrieval method based on esm1b\_t33\_650M\_UR50S\_L2 can retrieve positive samples with extremely low percent identity compared to BLAST. We consider that this phenomenon arises due to the nature of BLAST as a local alignment algorithm that relies on Markov models, essentially functioning as a greedy algorithm. Therefore, samples with high percent identity have a natural advantage in the ranking by BLAST. On the other hand, PLMs are capable of learning the underlying information of proteins to some extent and can avoid relying solely on Percent Identity as the criterion. This enables fair evaluation of both high and low Percent Identity samples in the retrieval process.

\subsubsection{EC number Hit Rate}\label{subsubsec3}

As illustrated in Table \ref{tab:mytable2}, we compare EC number hit rate of various methods and PLMs in framework.

In our evaluation, we examine established tools such as BLAST and Foldseek (TM-align), alongside a suite of cutting-edge PLMs: esm1b\_t33\_650M\_UR50S, esm2\_t33\_650M\_UR50D, esm2\_t36\_3B\_UR50D, esm\_msa1b\_t12\_100M\_UR50S, prot\_t5\_xl\_uniref50, and ProtGPT2. Complementing these, we assess protein structure models including esm\_if1\_gvp4\_t16\_142M\_UR50 and MIF-ST. For every PLM, we employ four distinct similarity measures—L2 distance, cosine similarity, inner product, and norm\_L2—to comprehensively gauge their retrieval performance.

\begin{figure}[ht]
    \centering
    \label{mypic:08}
    \includegraphics[scale=0.7]{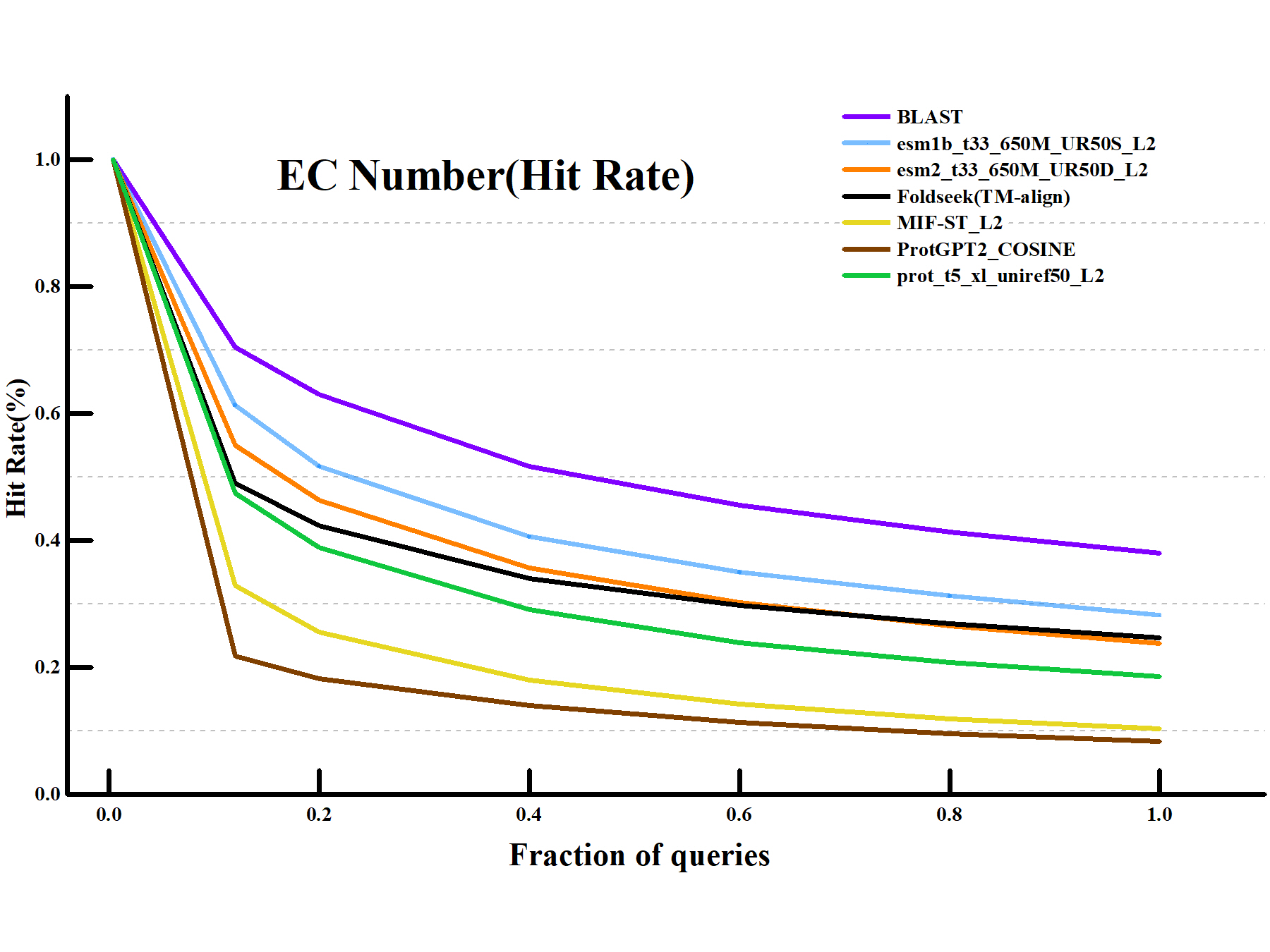}
    \caption{EC number hit rate.To mitigate visual redundancy, selectivity is exercised in showcasing results with representative value. Notably, for comparable PLMs, the L2 similarity metric parallels COSINE's performance concerning hit rate efficacy. Furthermore, sequential models consistently outperform structural ones, exemplified by the superiority of sequence-based models over Foldseek (TM-align), while the inverse holds for structural models.}
\end{figure}

\begin{table}[ht]
\centering
\caption{EC number hit rate}
\label{tab:mytable2}
\begin{tabular}{cccccccc}
\hline
Method/PLM                                        & Metric             & 30              & 50              & 100             & 150             & 200             & 250             \\ \hline
BLAST                                             & /                  & 0.7043          & 0.6298          & 0.5170          & 0.4557          & 0.4137          & 0.3800          \\ \hline
\multirow{4}{*}{\textbf{\makecell{esm1b\_t33\\\_650M\_UR50S}}} & \textbf{COSINE} & \textbf{0.6127} & \textbf{0.5170} & \textbf{0.4048} & \textbf{0.3486} & \textbf{0.3112} & \textbf{0.2801} \\
                                                  & IP                 & 0.2368          & 0.2127          & 0.1771          & 0.1563          & 0.1393          & 0.1263          \\
                                                  & \textbf{L2}        & \textbf{0.6132} & \textbf{0.5165} & \textbf{0.4059} & \textbf{0.3499} & \textbf{0.3127} & \textbf{0.2819} \\
                                                  & \textbf{norm\_L2}  & \textbf{0.6127} & \textbf{0.5170} & \textbf{0.4048} & \textbf{0.3486} & \textbf{0.3112} & \textbf{0.2801} \\ \hline
\multirow{4}{*}{\makecell{esm2\_t36\\\_3B\_UR50D}}             & COSINE          & 0.5952          & 0.5077          & 0.3952          & 0.3355          & 0.2975          & 0.2678          \\
                                                  & IP                 & 0.0192          & 0.0206          & 0.0191          & 0.0180          & 0.0176          & 0.0173          \\
                                                  & L2                 & 0.5851          & 0.4955          & 0.3842          & 0.3270          & 0.2897          & 0.2601          \\
                                                  & norm\_L2           & 0.5952          & 0.5079          & 0.3951          & 0.3355          & 0.2975          & 0.2678          \\ \hline
\multirow{4}{*}{\makecell{esm2\_t33\\\_650M\_UR50D}}           & COSINE          & 0.5558          & 0.4668          & 0.3595          & 0.3065          & 0.2698          & 0.2427          \\
                                                  & IP                 & 0.0124          & 0.0112          & 0.0114          & 0.0109          & 0.0111          & 0.0112          \\
                                                  & L2                 & 0.5500          & 0.4630          & 0.3563          & 0.3020          & 0.2652          & 0.2378          \\
                                                  & norm\_L2           & 0.5560          & 0.4671          & 0.3599          & 0.3069          & 0.2701          & 0.2429          \\ \hline
\multirow{4}{*}{\makecell{esm\_if1\_gvp4\\\_t16\_142M\_UR50}}  & COSINE          & 0.2567          & 0.1980          & 0.1325          & 0.1043          & 0.0885          & 0.0769          \\
                                                  & IP                 & 0.1390          & 0.1127          & 0.0808          & 0.0658          & 0.0561          & 0.0493          \\
                                                  & L2                 & 0.2469          & 0.1889          & 0.1274          & 0.1006          & 0.0851          & 0.0742          \\
                                                  & norm\_L2           & 0.2582          & 0.1992          & 0.1337          & 0.1050          & 0.0891          & 0.0775          \\ \hline
Foldseek                                          & /                  & 0.4897          & 0.4236          & 0.3403          & 0.2974          & 0.2686          & 0.2472          \\ \hline
\multirow{4}{*}{MIF-ST}                           & COSINE          & 0.3388          & 0.2658          & 0.1888          & 0.1506          & 0.1267          & 0.1106          \\
                                                  & IP                 & 0.0174          & 0.0186          & 0.0197          & 0.0193          & 0.0193          & 0.0193          \\
                                                  & L2                 & 0.3289          & 0.2560          & 0.1797          & 0.1423          & 0.1194          & 0.1036          \\
                                                  & norm\_L2           & 0.3392          & 0.2661          & 0.1887          & 0.1505          & 0.1266          & 0.1105          \\ \hline
\multirow{4}{*}{\makecell{esm\_msa1b\\\_t12\_100M\_UR50S}}     & COSINE          & 0.4398          & 0.3637          & 0.2772          & 0.2346          & 0.2080          & 0.1892          \\
                                                  & IP                 & 0.0676          & 0.0624          & 0.0540          & 0.0494          & 0.0463          & 0.0434          \\
                                                  & L2                 & 0.4428          & 0.3670          & 0.2791          & 0.2363          & 0.2101          & 0.1913          \\
                                                  & norm\_L2           & 0.4400          & 0.3638          & 0.2772          & 0.2346          & 0.2080          & 0.1893          \\ \hline
\multirow{4}{*}{ProtGPT2}                         & COSINE          & 0.2176          & 0.1829          & 0.1403          & 0.1135          & 0.0958          & 0.0838          \\
                                                  & IP                 & 0.0372          & 0.0332          & 0.0312          & 0.0298          & 0.0285          & 0.0273          \\
                                                  & L2                 & 0.2159          & 0.1804          & 0.1366          & 0.1093          & 0.0918          & 0.0800          \\
                                                  & norm\_L2           & 0.2168          & 0.1823          & 0.1400          & 0.1131          & 0.0954          & 0.0833          \\ \hline
\multirow{4}{*}{\makecell{prot\_t5\\\_xl\_uniref50}}           & COSINE          & 0.5752          & 0.4853          & 0.3761          & 0.3190          & 0.2817          & 0.2526          \\
                                                  & IP                 & 0.1576          & 0.1321          & 0.1012          & 0.0848          & 0.0748          & 0.0680          \\
                                                  & L2                 & 0.4741          & 0.3889          & 0.2911          & 0.2395          & 0.2082          & 0.1856          \\
                                                  & norm\_L2           & 0.5757          & 0.4850          & 0.3755          & 0.3184          & 0.2809          & 0.2519          \\ \hline
\end{tabular}
\end{table}

We select representative results for visualization. It can be observed that in this benchmark, sequence models generally outperform structural models. This could be due to slight discrepancies between the predicted protein tertiary structures by AlphaFold2 and the true structures. Among the sequence models, esm2\_t33\_650M\_UR50D performs slightly aboved average, while prot\_t5\_xl\_uniref50 and esm2\_t36\_3B\_UR50D, which have larger model sizes, demonstrate enhanced performance. The best-performing model is esm1b\_t33\_650M\_UR50S, which utilizes absolute position encoding. Additionally, in the structural models, MIF-ST generally outperform esm\_if1\_gvp4\_t16\_142M\_UR50. Among the four similarity metrics, L2 distance, cosine similarity, and normalized L2 distance (norm\_L2) show similar performance, all significantly higher than the inner product (IP).

\subsubsection{TP hits up to the 1st FP}\label{subsubsec4}

In the evaluation of retrieval methods, stability is also a key metric. In this study, we examine the number of true positive samples before encountering the first false positive sample in the retrieval results of the aforementioned methods. A higher number indicates greater stability of the retrieval method. Please refer to Table \ref{tab:mytable3} for details.

\begin{table}[ht]
\centering
\caption{TP hits up to 1st FP}
\label{tab:mytable3}
\begin{tabular}{ccccc}
\hline
                                 & COSINE        & IP              & L2               & norm\_L2          \\ \hline
BLAST                            & \multicolumn{4}{c}{45.39315353}                                           \\
\textbf{esm1b\_t33\_650M\_UR50S} & \textbf{31.8485} & \textbf{9.0000} & \textbf{31.7189} & \textbf{31.8423}  \\
\textbf{esm2\_t36\_3B\_UR50D}    & \textbf{34.6110} & \textbf{1.0197} & \textbf{32.7593} & \textbf{34.62137} \\
esm2\_t33\_650M\_UR50D           & 28.6203          & 1.0197          & 28.0207          & 28.6151           \\
esm\_if1\_gvp4\_t16\_142M\_UR50  & 5.5913           & 2.6390          & 5.4492           & 5.6203            \\
\textbf{Foldseek}                & \multicolumn{4}{c}{\textbf{65.84958506}}                                  \\
MIF-ST                           & 8.5726           & 1.0726          & 8.1950           & 8.5705            \\
esm\_msa1b\_t12\_100M\_UR50S     & 18.9367          & 2.3133          & 19.5954          & 18.9305           \\
ProtGPT2                         & 9.1359           & 1.4243          & 8.7459           & 9.1608            \\
prot\_t5\_xl\_uniref50           & 30.28423237      & 4.9824          & 19.6473          & 30.22303          \\ \hline
\end{tabular}
\end{table}

The experimental results show that esm2\_t36\_3B\_UR50D has the best stability among the PLMs, with a true positive (TP) hits up to the first false positive (1st FP) value as high as 34.62136929. Next are esm1b\_t33\_650M\_UR50S and prot\_t5\_xl\_uniref50, which have similar performances, but it is worth noting that the TP hits up to 1st FP for prot\_t5\_xl\_uniref50\_L2 are much lower than prot\_t5\_xl\_uniref50\_COSINE and prot\_t5\_xl\_uniref50\_norm\_L2. The best-performing PLMs in terms of Hit Rate is esm1b\_t33\_650M\_UR50S, with a slightly lower TP hits up to the 1st FP value compared to esm1b\_t33\_650M\_UR50S and prot\_t5\_xl\_uniref50. Among the four similarity metrics, cosine similarity and normalized L2 distance (norm\_L2) exhibit similar performance, both significantly outperforming L2 distance and inner product (IP).

\begin{figure}[htbp]
    \centering
    \label{mypic:09}
    \includegraphics[scale=0.5]{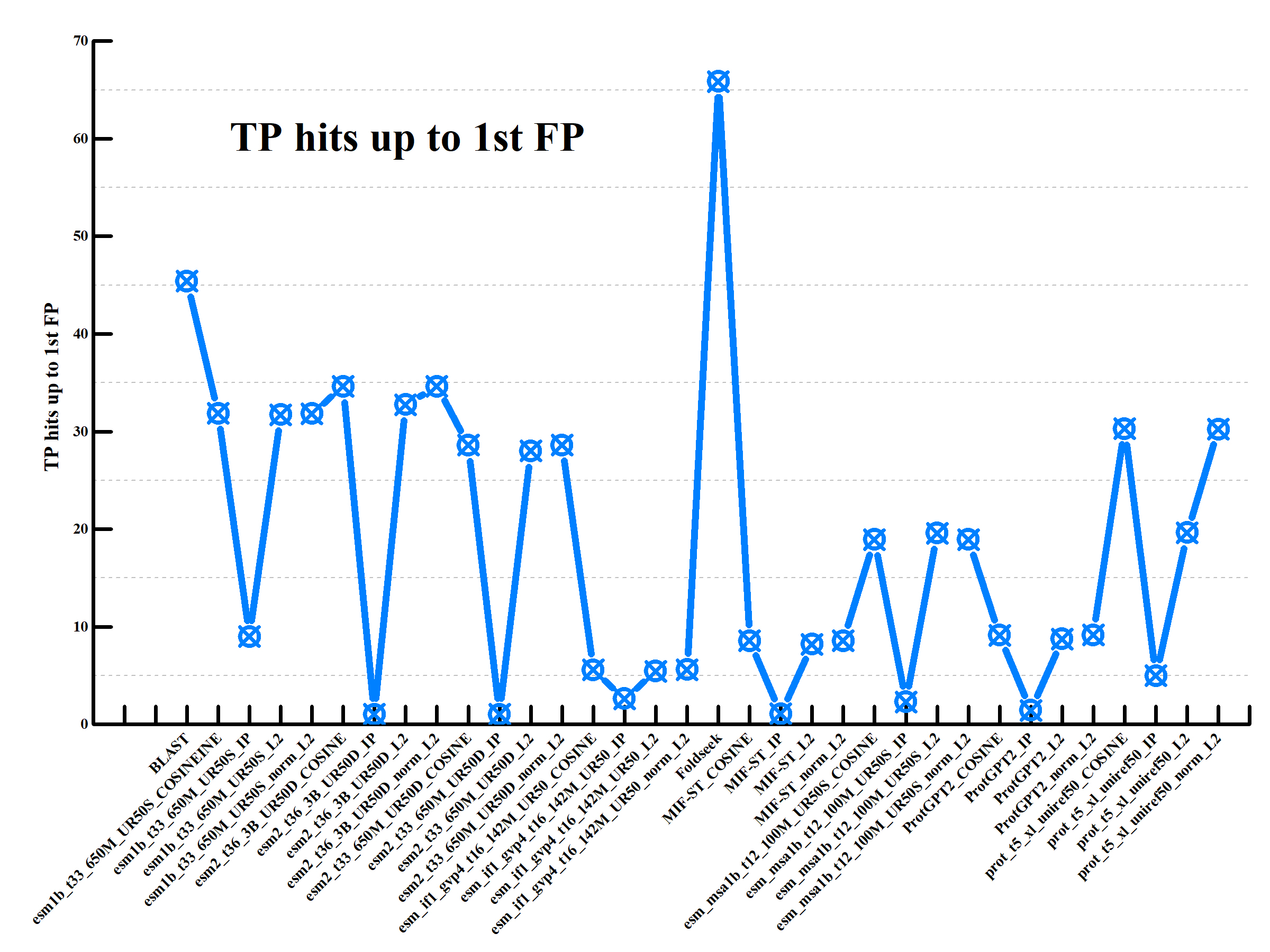}
    \caption{TP hits up to 1st FP}
\end{figure} 

\subsubsection{Further validation}\label{subsubsec6}

As the size of protein databases increases, BLAST would find it increasingly challenging to identify positive samples with low percent identity. On the other hand, PLMs could uncover the underlying information and tertiary structure of proteins, enabling them to surpass local optima sets characterized by percent identity. In this study, we construct a vector database based on the UniProtKB/TrEMBL, which contains 251,131,639 protein entries. We attempt to reproduce the process of finding pRed-14 and pRed-15, as described by Christopher R. B. Swanson and Grayson J. Ford in their paper "Biocatalytic Cascades toward Iminosugar Scaffolds Reveal Promiscuous Activity of Shikimate Dehydrogenases."\cite{bib9}

In their paper, Swanson and Ford first conducted a screening test for the oxidative direction of amine targets using a high-throughput colorimetric method. They identified 384 intragenomic imine reductases (IReds). The search required the results to include two segments of the given gene cluster. Finally, they selected a group of 24 potential enzymes (pRed-1 to pRed-24) with diverse classifications and sequence homology. Multiple sequence alignments revealed that many of the identified "reductases" did not exhibit conserved residues with the 384 internal IRed gene clusters. The percent identity between pRed-15 and the others with the same function, PtDH, AroE, and YdiB, was as low as 16.14\%.

\begin{figure}[htbp]
    \centering

    \includegraphics[scale=0.5]{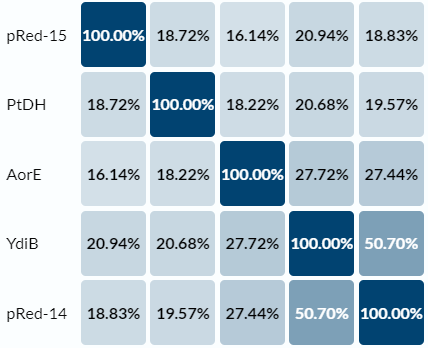}
    \caption{Alignment of PtDH, AroE, YdiB, pRed-14 and pRed-15}
    \label{mypic:1119}
\end{figure}

In this study, we start the retrieval from PtDH, AroE, and YdiB (seed proteins in paper\cite{bib9}). As shown in Figure\ref{mypic:1119}, except for YdiB and pRed-14, the percentage identity is below 30\%. Both BLAST and framework with esm1b\_t33\_650M\_UR50S and L2 similarity are tested. The threshold is adjusted to encompass a broader retrieval result until one method first identifies pRed-14 and pRed-15. The methods that prioritized finding the expected targets are documented. The results are shown in the Table below.

\begin{table}[htbp]
\centering
\caption{Framework and BLAST on pRed-14 \& pRed-15}\label{tab:mytable5}%
\begin{tabular}{ccccc}
\toprule
\textbf{}        & \textbf{Framework} & \textbf{BLAST} & \textbf{Framework} & \textbf{BLAST} \\
\hline
\textbf{}        & \multicolumn{2}{c}{\textbf{pRed-14}}                  & \multicolumn{2}{c}{\textbf{pRed-15}}                  \\ 
\midrule
\textbf{PtDH}  & \textbf{\checkmark}                           & \textbf{}      & \textbf{\checkmark}                           & \textbf{}      \\
\textbf{AroE}  & \textbf{}                           & \textbf{\checkmark}      & \textbf{\checkmark}                           & \textbf{}      \\
\textbf{YdiB} & \textbf{}                           & \textbf{\checkmark}      & \textbf{\checkmark}                           & \textbf{}      \\
\bottomrule
\end{tabular}
\end{table}

It could be observed that BLAST can more rapidly identify pRed-14 when initiating the search from AroE and YdiB due to their high percent identity. Surprisingly, even though PtDH has a modest 19.57\% identity with pRed-14, our framework manages to identify the target sooner. In the case of retrieving pRed-15, our framework exhibits comprehensive superiority compared to BLAST. The results indicates that the retrieval framework outperformd BLAST in retrieving sequences with low percent identity. This approach avoid the pitfall of relying solely on percent identity during retrieval, allowing for a fair evaluation of all samples.

\section{Conclusion}\label{sec5}

In this study, we present an innovative PLMs based protein retrieval framework, accompanied by a rigorous benchmark for evaluating its performance and the efficacy of PLMs in retrieval tasks. Our investigation encompasses a wide array of prominent sequence-based and structure-based PLMs, with ablation studies shedding light on the utility of four embedding similarity metrics.

We juxtapose the retrieval outcomes of PLM-based methods against those from established tools such as BLAST and Foldseek employing TM-align, meticulously examining their disparities. Our findings highlight that BLAST's high hit rate owes to its algorithm's compatibility with protein sequence regularities, where similar sequences frequently denote similar structures and functions. However, in biological contexts, specific functions can be conferred by localized regions, as illustrated by Swanson and Ford's work on pRed-14 and pRed-15, necessitating retrieval approaches that can discern functionally relevant local similarities without being misled by overall sequence identity. Addressing these complexities is vital for unbiased and effective retrieval. Our PLMs based retrieval framework is shown to partially capture proteins' latent information through deep learning, enabling retrieval guided by a more profound protein understanding. This constitutes a fresh paradigm for advancing protein retrieval methodologies.

Despite demonstrating enhanced protein comprehension, our PLM retrieval framework exhibits variable hit rates and stability compared to conventional methods. In-depth analysis reveals that while BLAST excels with probes featuring extensive multiple sequence alignments (MSAs), our framework outperforms BLAST with niche probes lacking extensive homologous data. This prompts our hypothesis that fine-tuning or training PLMs on targeted family or MSA data could significantly boost retrieval accuracy and stability. Relative to Foldseek (TM-align), our framework achieves higher hit rates, albeit with room for enhancing stability. Our work introduces a potent new vantage point in protein retrieval research, stimulating further exploration towards developing high-performance and reliable retrieval methods, and fostering advancements in PLM research and development.

\newpage

\bibliographystyle{unsrt}  
\bibliography{main}

\end{document}